\documentclass[runningheads]{llncs}
\usepackage{graphicx}
\usepackage{diagbox}
\usepackage{color,soul}
\usepackage[misc]{ifsym}
\usepackage{graphicx}
\usepackage{verbatim}
\begin{document}

\title{Improving Lesion Segmentation for Diabetic Retinopathy using Adversarial Learning}
%
%
\author{Qiqi Xiao\inst{1}\orcidID{0000-0002-9695-0603} \and
Jiaxu Zou\inst{1}\orcidID{0000-0003-4861-8072} \and
Muqiao Yang\inst{1}\orcidID{0000-0001-6273-0138} \and
Alex Gaudio\inst{1}\orcidID{0000-0003-1380-6620} \and
Kris Kitani\inst{1}\orcidID{0000-0002-9389-4060} \and
Asim Smailagic\inst{1}\textsuperscript{(\Letter)}\orcidID{0000-0001-8524-997X} \and
Pedro Costa\inst{2}\orcidID{0000-0002-9528-1292} \and
Min Xu\inst{1}\orcidID{0000-0002-0881-5891}}
\authorrunning{Q. Xiao et al.}
\institute{Carnegie Mellon University, Pittsburgh PA 15213, USA \\
\email{\{qiqix,jiaxuz,agaudio\}@andrew.cmu.edu, \{muqiaoy,kkitani,asim,mxu1\}@cs.cmu.edu} \and INESC TEC, Porto, Portugal \\ \email{pedro.vendascosta@gmail.com}}
\maketitle              
\begin{abstract}
Diabetic Retinopathy (DR) is a leading cause of blindness in working age adults. DR lesions can be challenging to identify in fundus images, and automatic DR detection systems can offer strong clinical value. Of the publicly available labeled datasets for DR, the Indian Diabetic Retinopathy Image Dataset (IDRiD) presents retinal fundus images with pixel-level annotations of four distinct lesions: microaneurysms, hemorrhages, soft exudates and hard exudates. We utilize the HEDNet edge detector to solve a semantic segmentation task on this dataset, and then propose an end-to-end system for pixel-level segmentation of DR lesions by incorporating HEDNet into a Conditional Generative Adversarial Network (cGAN).  We design a loss function that adds adversarial loss to segmentation loss. Our experiments show that the addition of the adversarial loss improves the lesion segmentation performance over the baseline.

\end{abstract}
\section{Introduction}
Diabetic Retinopathy (DR) is an eye disease caused by damage to the retinal blood vessels of diabetic patients. Since the disease is relatively asymptomatic until the patient experiences loss of vision, physicians recommend regular screenings for diabetic patients. Analysis of high resolution fundus images obtained during the screening requires considerable time and effort by trained clinicians, as lesions can be hard to detect.

While the diagnosis of the disease ultimately requires a physician,  automated detection of DR lesions can improve patient outcomes.  Recent developments in machine learning and computer vision that enable accurate classification and localization are well suited to the DR detection task.  Of particular interest are pixel level annotations of DR lesions that suggest to physicians where in the image the lesions should be. Automated detection methods save time and can reduce uncertainty in DR diagnosis.

The datasets available for DR strongly influence development of automated detection algorithms.  Publicly available datasets for DR, such as Messidor \cite{dset_messidor}, DRIVE \cite{dset_drive}, STARE \cite{dset_stare} and DIARETDB \cite{dset_diaret}, contain annotations of the whole image or of sub-regions of the image.  Unfortunately, detection algorithms built from these datasets tend to make image level or patch level predictions, which by design has limited utility to a clinician who needs to explain the underlying factors leading to the diagnosis.  A system capable of accurate pixel-level segmentation is more explainable and provides better value to clinicians. 

In this work, we use the Indian Diabetic Retinopathy Image Dataset (IDRiD) \cite{idrid}. To the best of our knowledge, IDRiD is the first public database for DR containing pixel level annotations of four typical DR lesions: microaneurysms (MA), hemorrhages (HE), hard exudates (EX), and soft exudates (SE).  Physicians assess combinations of these lesions to diagnose various grades of DR. 

\begin{figure}
\centering
\includegraphics[]{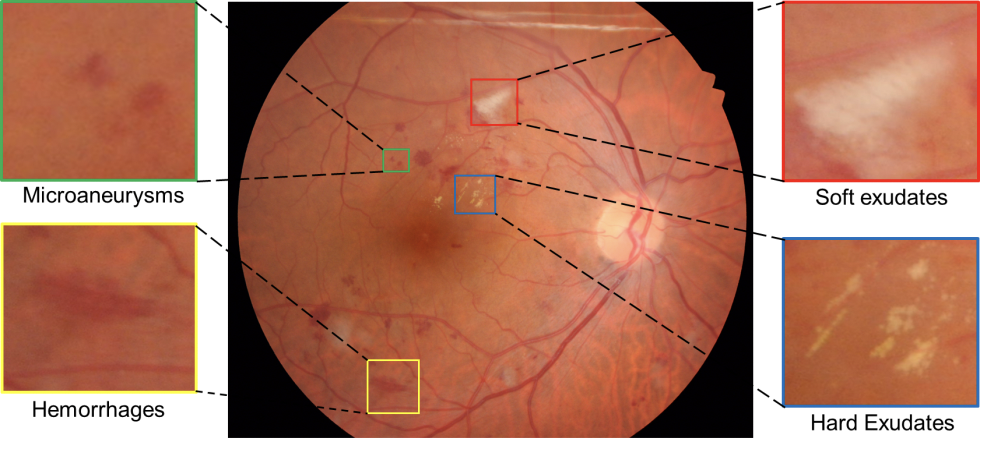}
\caption{Color fundus photograph containing different retinal lesions associated with diabetic retinopathy. Enlarged parts illustrating presence of Microaneurysms, Soft Exudates, Hemorrhages and Hard Exudates.} \label{fig1}
\end{figure} 

Our method uses the Holistically-Nested Edge Detection (HEDNet) network \cite{hednet} to compute a segmentation map from a fundus image.  To enhance HEDNet segmentation performance, we incorporate this model into a conditional Generative Adversarial Network (GAN) with a standard PatchGAN discriminator. Our method is end-to-end, and we show that the addition of adversarial loss can improve the lesion segmentation performance of diabetic retinopathy images.

\section{Related Work}

\subsection{HEDNet in Semantic Segmentation}

Semantic Segmentation is an image-to-image translation method that aims to identify regions and structures in an input image. These methods solve pixel-level classification problems, where the classes are pre-defined. For example, semantic segmentation of street view images produces classes like person, vehicle, building, etc. The result of such segmentation is fine-grained and thus contains more information about the scene than both simple classification and bounding box detection.  In the context of the IDRiD dataset for DR, each pixel can be annotated as one of four lesion types or healthy.

Holistically-Nested Edge Detection (HEDNet) \cite{hednet} is a state-of-art algorithm proposed to solve image-to-image problems with a deep convolutional neural network.  Unlike traditional edge detectors, HEDNet can generate semantically meaningful edge maps that identify object contours. Experiments on Berkeley Segmentation Dataset show that HEDNet performs much better than traditional edge detection algorithms like Canny edge detection, and it also outperforms patch-based edge detection algorithms in terms of speed and accuracy \cite{hednet}. Although it is originally proposed to solve edge detection for natural images, we show that HEDNet is capable of solving the segmentation problem as well.

Considering the effectiveness of HEDNet for edge detection and semantically meaningful contour maps, we choose to base our work on top of this architecture.  We show that HEDNet is capable of solving the segmentation problem on the IDRiD dataset.

\subsection{GAN in Semantic Segmentation}

Classification algorithms, such as those for semantic segmentation, perform well when the task has a clearly defined objective.  In practice, however, the objective function used often incorporates hidden assumptions that can be overly simplistic.  For instance, the classification setting might assume that each pixel belongs to precisely one class, but in reality, a pixel could represent presence of both soft exudates (which occur in the Nerve Fiber Layer of the Retina) and hard exudates (which occur deeper in the retina).  When we think about labeling these pixels, the multi-class setting breaks down.  Therefore, in semantic segmentation tasks, the objective can be challenging to define because we need to consider all possible assumptions and we may not know in advance what they are.

Semantic segmentation tasks have been framed as adversarial generative modeling problems (\cite{adversarial_semantic_segmentation}, \cite{adversarial_semi_supervised_semantic_segmentation}), where the generative model's objective function is learned.  For instance, in the area of medical image processing, Splenomegaly Segmentation Network(SSNet) \cite{ssnet} utilizes conditional generative adversarial networks (cGAN) \cite{cgan} to solve the spleen volume estimation problem, and the work shows significant improvement over the baseline on a medical dataset containing 60 MRI images.

While we use HEDNet to solve a straight-forward classification problem with the pixel-wise ground truth labels from IDRiD, we can also evaluate how realistic the HEDNet annotations are.  Therefore, we present semantic segmentation on IDRiD as a generative modeling task, and we train HEDNet to both classify pixels correctly and generate realistic segmentation maps of typical DR lesions.

\section{Methodology}
\begin{figure}
\centering
\includegraphics[scale=0.4]{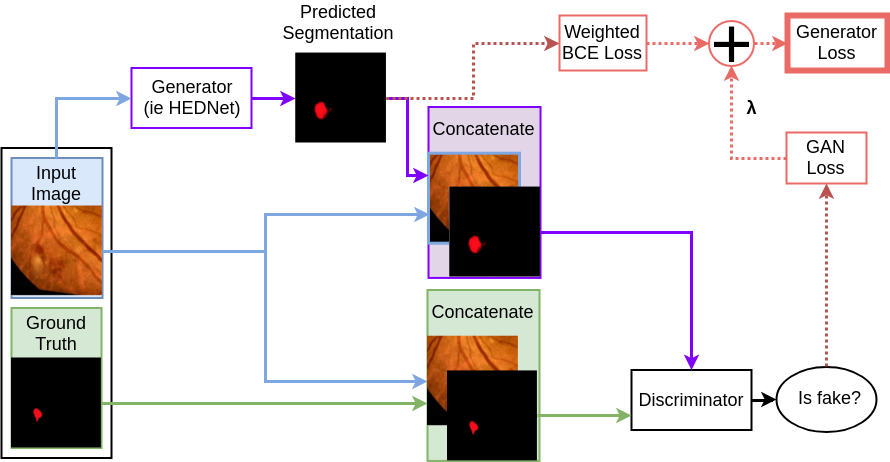}
\caption{Main framework of conditional generative adversarial network} \label{fig2}
\end{figure}

In this section we start by showing how we preprocess the retinal images, then we explain how we use an image-to-image network to segment DR lesions and how we combine the GAN loss to further refine the segmentation results. An overview of our model structure is shown in Figure \ref{fig2}.

\subsection{Preprocessing Steps}

Before we feed the raw images into our network, we consider using illumination correction and contrast enhancement techniques on the retinal images for better image enhancement.  \\

\subsubsection{Brightness Balance}
Since the dataset is sampled from different lesions and tissues, there might exist some inconsistency in the brightness of the whole dataset. To avoid the imbalance of brightness among different images, we force each training and test image to have an average pixel intensity equal to the average pixel intensity of the training set.

\subsubsection{Contrast Enhancement}
Contrast enhancement ensures the pixel intensities cover a wide range of values, which can make details more readily apparent.  We applied CLAHE (Contrast Limited Adaptive Histogram Equalization) for contrast enhancement. CLAHE affects small regions of the image instead of the entire image \cite{clahe}, and it can have significantly better performance than the regular histogram equalization.

\subsubsection{Denoising}
In practical situations, intrinsic and extrinsic conditions related to capture of the image result in different kinds of noise, and denoising is a fundamental challenge in image processing. Here we assume that the images contain Gaussian white noise, and apply the Non-local Means Denoising algorithm \cite{denoise}. Additionally, we apply a bilateral filter the the image, which replaces the intensity of each pixel with a weighted average of intensity values from nearby pixels, thus preserving edge information while the noise is minimized.

\subsection{Image to Image Network}

The image to image network for segmentation we use is HEDNet \cite{hednet}. HEDNet builds on VGGNet \cite{vgg} by adding side outputs to the last convolutional layer in each stage and by removing the last stage and all fully connected layers. The VGGNet structure is initialized with weights pre-trained on ImageNet and then fine tuned. The side outputs are fused together via a trainable weighted-fusion layer; since each output corresponds to a different stage of VGGNet, the fusion enables a multi-scale representation of the output.  This fully convolutional architecture allows HEDNet to maintain both high level information and low level details. \\

All side outputs from HEDNet are concatenated, and therefore each stage of the network contributes to the final pixel-wise binary cross-entropy (BCE) loss. This is known as deep supervision in the sense that each stage can be interpreted as an individual network output solving the learning task at a specific scales.  Experiments have shown that with only the fusion layer loss, a large amount of edge information is lost on high level side outputs. \\

Diabetic Retinopathy lesions typically make up a very small proportion of a diseased fundus image and do not exist for images of healthy eyes.  As a result, the ground truth is highly imbalanced. We use a class-balancing weight $\beta$, which differentiates cross-entropy loss for positive and negative samples: \\
\begin{equation}
    Loss_{weight\_BCE}=-(\beta\cdot y\textrm{log}p+(1-y)\textrm{log}(1-p))
\end{equation}
where $y$ is the binary indicator 0 or 1 and $p$ is the predicted probability of observation of the positive class. \\


\subsection{GAN Loss}

Inspired by Splennomegaly Segmentation Network (SSNet) \cite{ssnet}, we find that the variations in both size and shape of the lesions from the Diabetic Retinopathy can introduce a large number of false positive and false negative labelings, and conditional GAN \cite{cgan} is an effective approach to improve generalization ability. The generator of SSNet is a Global Convolutional Network (GCN), which is inherently an image-to-image fully convolutional network with a large receptive field. Since we want to output an image of the same size with the input image, here we need an equivalent kernel size, therefore we propose to use HEDNet to replace the GCN for diabetic retinopathy. \\

Furthermore, a conditional GAN is used to discriminate the output, whose architecture is the same as infoGAN \cite{infogan}. The discriminator utilizes the framework of PatchGAN \cite{patchgan}, where the input image is split into smaller patches, and each small image patch is applied with a cross entropy loss to decide whether that patch is fake or real. 
The input to the discriminator is the concatenation of the original image patch and the corresponding segmentation output from the generator, which is a 4-channel tensor. Therefore, we can see that the discriminator learns the joint distribution of the input and the segmentation map conditioned on the input. \\


\subsection{Loss Function}

The final generator loss term is a weighted average of binary cross-entropy loss and GAN loss:
\begin{equation}
    Loss_{generator} = Loss_{weighted\_BCE} + \lambda\cdot Loss_{GAN}
\end{equation}
thus the goal of the network is to produce good segmentation with respect to ground truth as well as to make segmentation consistent such that the segmentation result seems real to the discriminator. Therefore, it is used to further refine the segmentation results. \\

\section{Experiments}

\subsection{Datasets}
We use the dataset from IDRiD challenge \cite{idrid}. This sub-challenge can be divided in four different tasks, which are lesion segmentation of Microaneurysms (MA), Soft Exudates (SE), Hard Exudates (EX) and Hemorrhages (HE).  Of the 54 training set images and 27 test set images, not all images contain all four lesion types.  Table \ref{tab1} shows the percent of images in the train and test sets respectively assigned to each of the four classes. We randomly divided the training set into training set with 80\% images and validation set with 20\% images.  Each image has resolution of $4288 \times 2848$. \\

\begin{table}[htb]
    \centering
    \begin{tabular}{|m{3cm}|c|c|p{0.2\columnwidth}|}
        \hline
        \diagbox[innerwidth=3cm]{Type}{Dataset}  & Training Set Images & Testing Set Images & Total Pathological Images\\
        \hline
        microaneurysms (MA) & 54 (100\%) & 27 (100\%) & 81 (100\%)\\
        \hline
        soft exudates (SE) & 26 (48\%)  & 14 (52\%) & 40 (49\%)\\
        \hline
        hard exudates (EX) & 54 (100\%) & 27 (100\%) & 81 (100\%)\\
        \hline
        hemorrhages (HE) & 53 (98\%) & 27 (100\%) & 80 (99\%)\\
        \hline 
    \end{tabular}
    \caption{Structure of IDRiD dataset.  Percentages show amount of images containing the given lesion type.} \label{tab1}
\end{table}

\subsection{Implementation Details}
\subsubsection{Hyperparameters}
We use a pixel value in [0, 1] for each lesion image and ground truth segmentation. We use a patch size of 128 for the SE, EX and HE models and patch size of 64 for the MA model. We set the weight $\beta$ in BCE loss to 10 to balance the positive and negative labels. We set the weight of the GAN loss $\lambda = 0.01$, as in SSNet. We use SGD as our optimizer for both HEDNet and the discriminator with an initial learning rate of 0.001 in both cases. For HEDNet, we decay the learning rate by 10\% every 200 epochs. For the discriminator model, we decay the learning rate by 10\% every 100 epochs. The momentum factor of the optimizer is 0.9. In addition, we apply L-2 weight decay with a rate of 0.0005. The training and validation batch size is 4 and the testing batch size is 1. For all experiments, the model is trained for 5000 epochs.

\subsubsection{Preprocessing}
For contrast enhancement, we apply the CLAHE technique with tiles of 8 $\times$ 8 pixels and a default contrast limit of 40. For denoising, we apply the Non-local Means Denoising algorithm with a filter strength of 10. We also normalize each channel of the lesion image to a mean of 0.485, 0.456, 0.406 and standard deviation of 0.229, 0.224, 0.225.

\subsubsection{Data Augmentation}
During training, we randomly crop each image to 512 $\times$ 512 pixels, and randomly rotate each image using a maximum angle of $20^{\circ}$.

\subsection{Performance Evaluation Metrics}

We use several performance metrics for evaluation, including Average Precision Score (AP), F-1 score and Precision-Recall Curve (PRC). All the 3 metrics reflect the precision and recall performance of the model on binary segmentation tasks from different perspectives. We compute the score of AP and F-1, and plot the PRC for each model on all the entire testing set. The evaluation is implemented using scikit-learn functions.

\subsection{Experimental results}

\subsubsection{Quantitative results}

We compare three models: UNET, HEDNet and HEDNet with cGAN.  UNET, which was originally proposed for biomedical image segmentation \cite{unet}, is a standard and widely used model \cite{unetpp,unetdense,unetcascade}.   Results show that our HEDNet+cGAN model improves over both HEDNet and UNET.  HEDNet+CGAN improves average precision of SE, HE and EX segmentations, and it improves the F1 score for MA, SE and HE. The average precision and F-1 scores are shown in Table \ref{tab2} and Table \ref{tab3} respectively. The results show that the model performs best on hard exudates, where it achieves the highest scores for both AP and F-1 score.  This can be explained by the pathological features of EX lesions. Hard Exudates are small shiny white or yellowish white deposits deep to the retinal vessels with sharp margins, which leads to high contrast in the images. We do not see an obvious improvement on MA segmentation from the experimental results, which is also related to pathological features of MA. Compared to other 3 types of lesions, microaneurysms are very small, lower contrast and share higher similarity to blood vessels, which can confuse the model to certain extent. The nearly consistent improvements of the HEDNet+cGAN model over the HEDNet model on all 4 lesion types for both evaluation metrics demonstrates the model strength under the cGAN framework. \\

\begin{table}[htb]
    \centering
    \begin{tabular}{|c|c|c|c|}
        \hline
        \diagbox[]{Model}{AP}  & UNET & HEDNet & HEDNet + cGAN \\
        \hline
        microaneurysms (MA) & 41.84\% & \textbf{44.03\%} & 43.92\% \\
        \hline
        soft exudates (SE) & 42.22\% & 43.07\% & \textbf{48.39\%} \\
        \hline
        hard exudates (EX) & 79.05\% & 83.98\% & \textbf{84.05\%} \\
        \hline
        hemorrhages (HE) & 41.93\% & 45.69\% & \textbf{48.12\%} \\
        \hline 
    \end{tabular}
    \caption{Average Precision on the test dataset on four lesions} \label{tab2}
\end{table}

\begin{table}[htb]
    \centering
    \begin{tabular}{|c|c|c|c|}
        \hline
        \diagbox[]{Model}{F-1 score}  & UNET & HEDNet & HEDNet + cGAN\\
        \hline
        microaneurysms (MA) & 41.76\% & 39.81\% & \textbf{42.98\%} \\
        \hline
        soft exudates (SE) & 27.88\% & 40.12\% & \textbf{43.98\%} \\
        \hline
        hard exudates (EX) & \textbf{69.90\%} & 68.94\% & 69.08\% \\
        \hline
        hemorrhages (HE) & 44.97\% & 45.00\% & \textbf{45.76\%} \\
        \hline 
    \end{tabular}
    \caption{F-1 Score on the test dataset on four lesions} \label{tab3}
\end{table}



\begin{figure}
\centering
\includegraphics[scale=0.35]{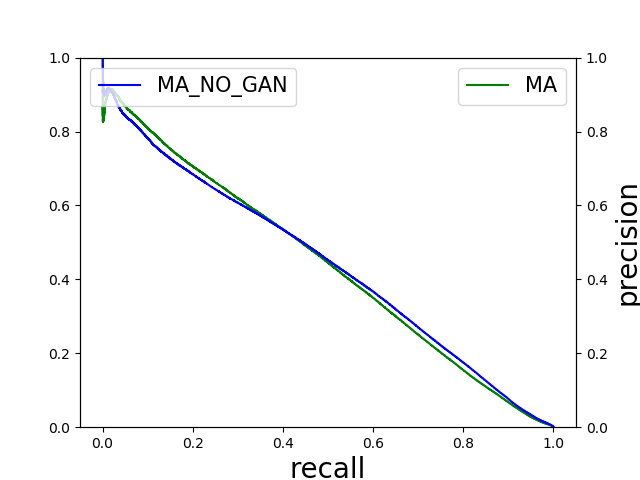}
\includegraphics[scale=0.35]{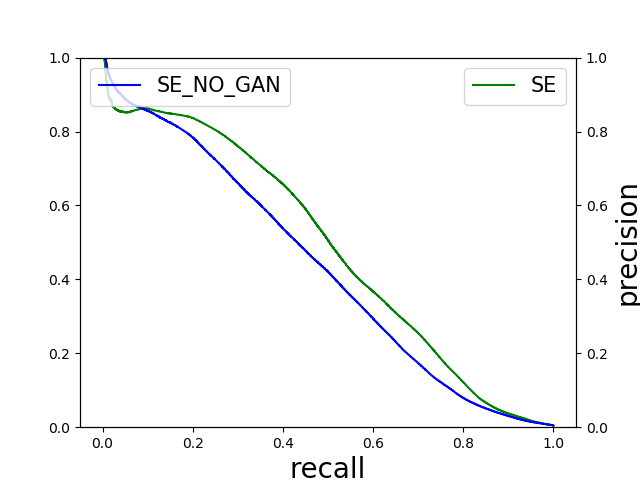}
\includegraphics[scale=0.35]{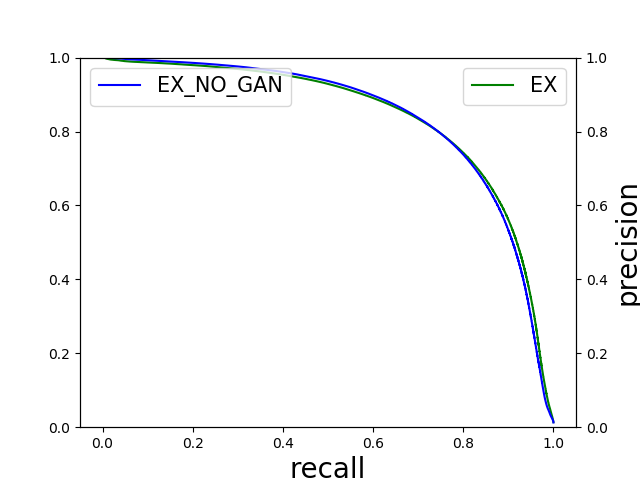}
\includegraphics[scale=0.35]{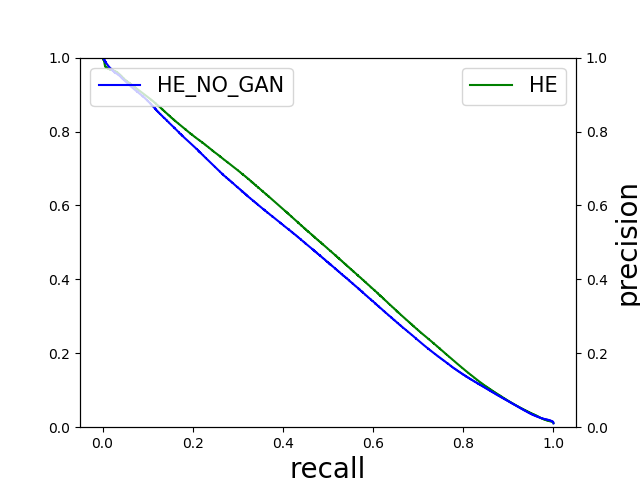}
\caption{Precision-Recall Curves for four models} \label{prc}
\end{figure}

\subsubsection{Qualitative Results}
\begin{figure}
\centering
\includegraphics[scale=0.256]{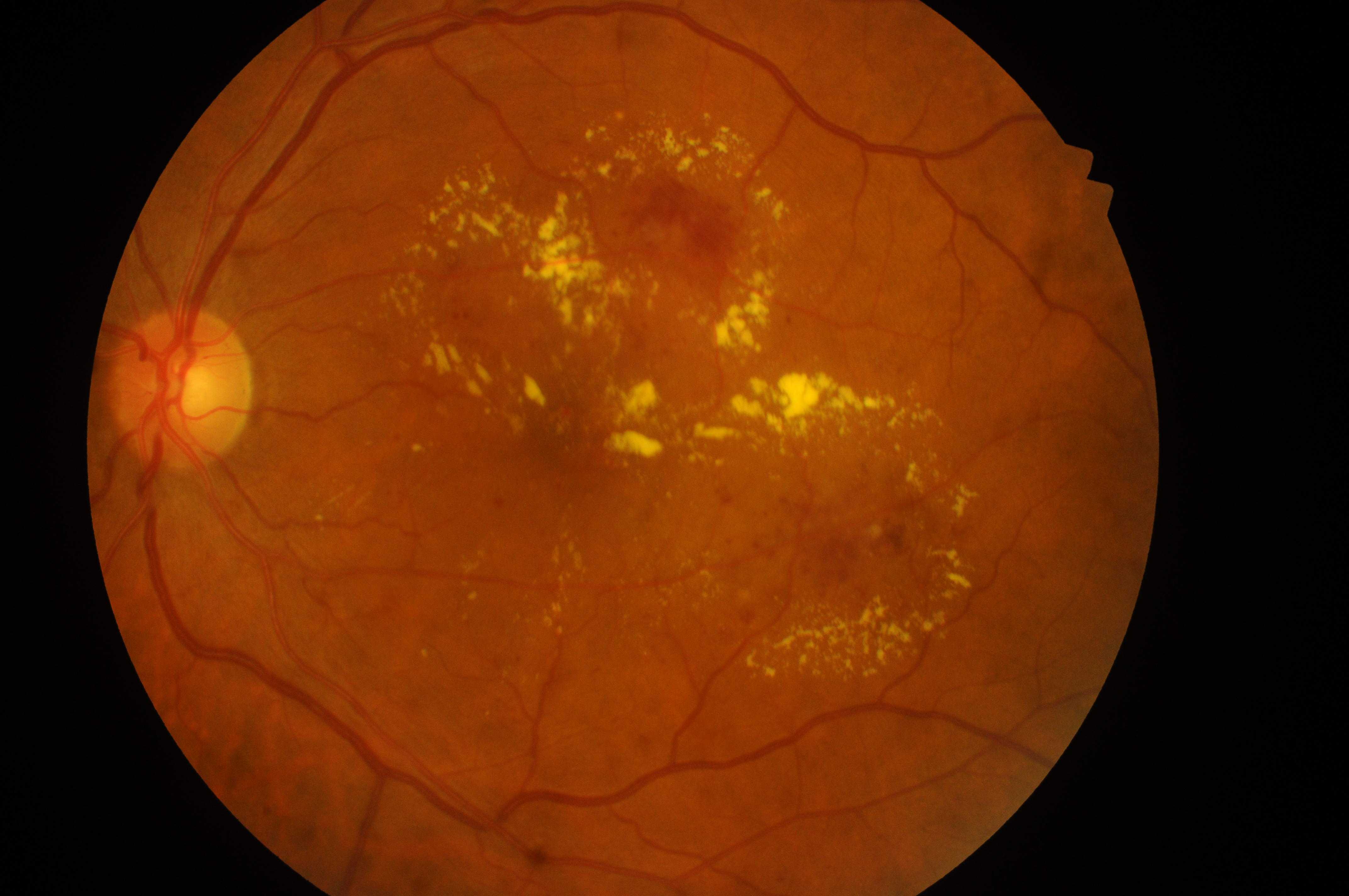}
\includegraphics[scale=0.02]{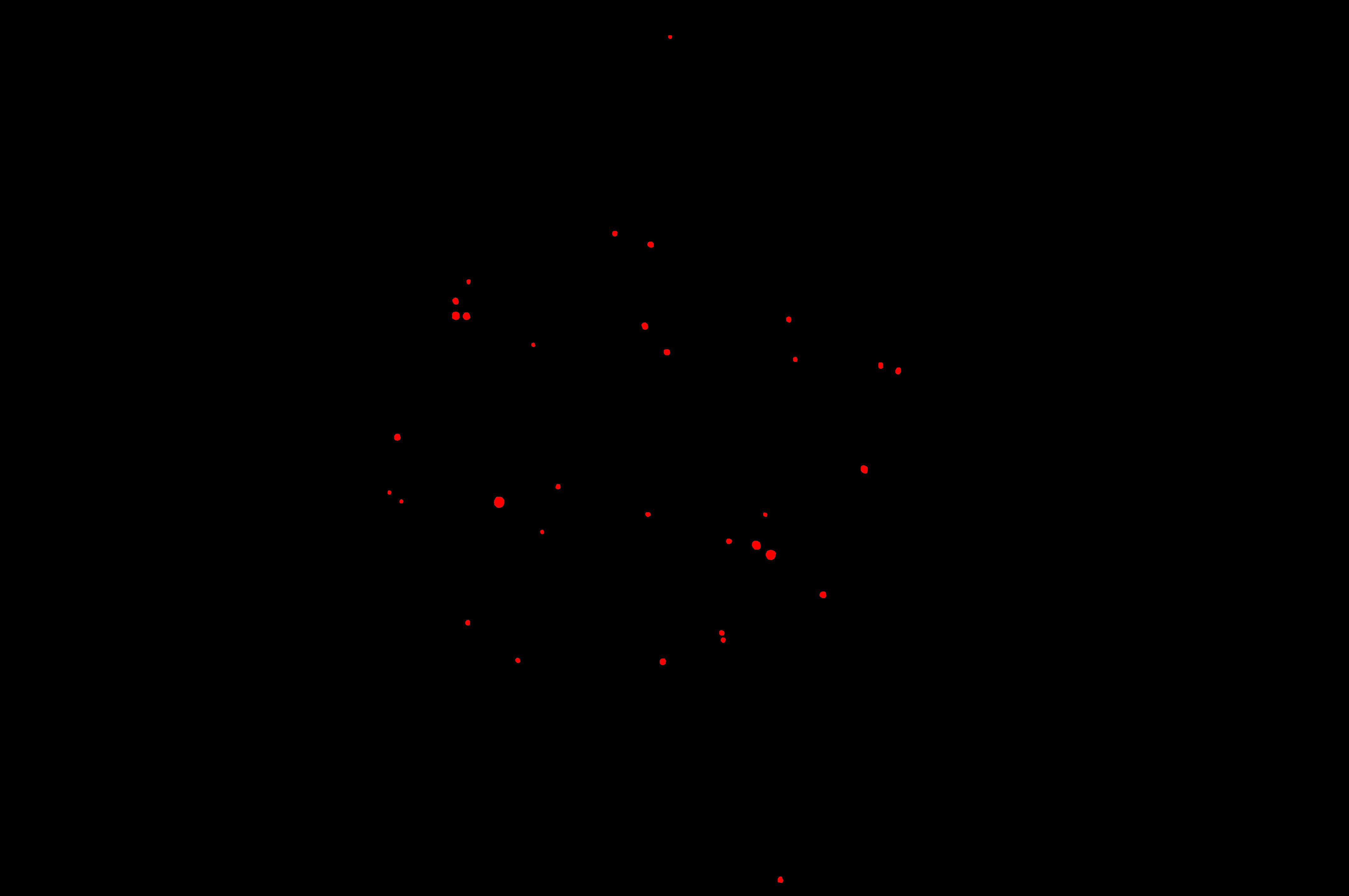}
\includegraphics[scale=0.02]{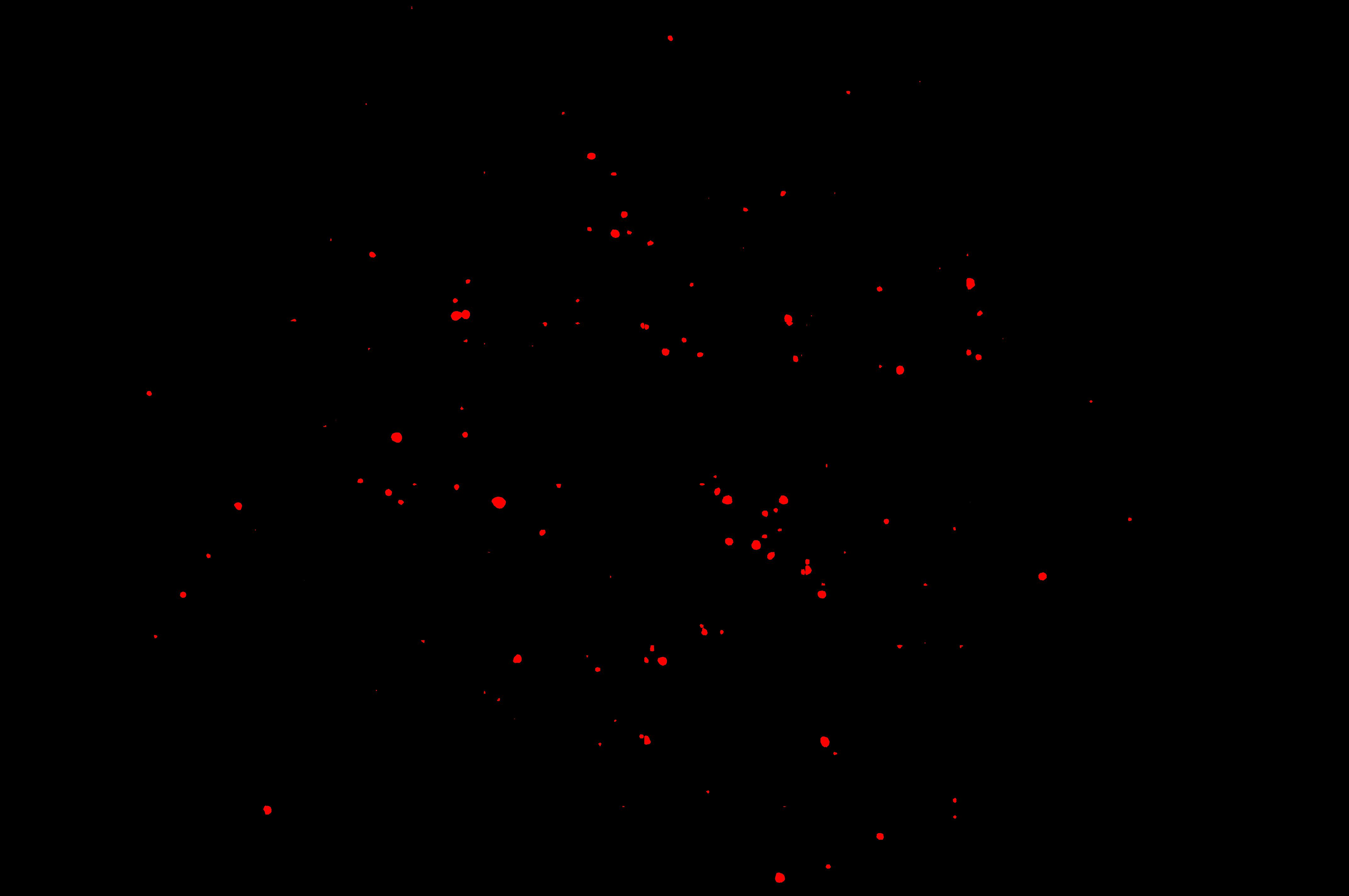}
\includegraphics[scale=0.02]{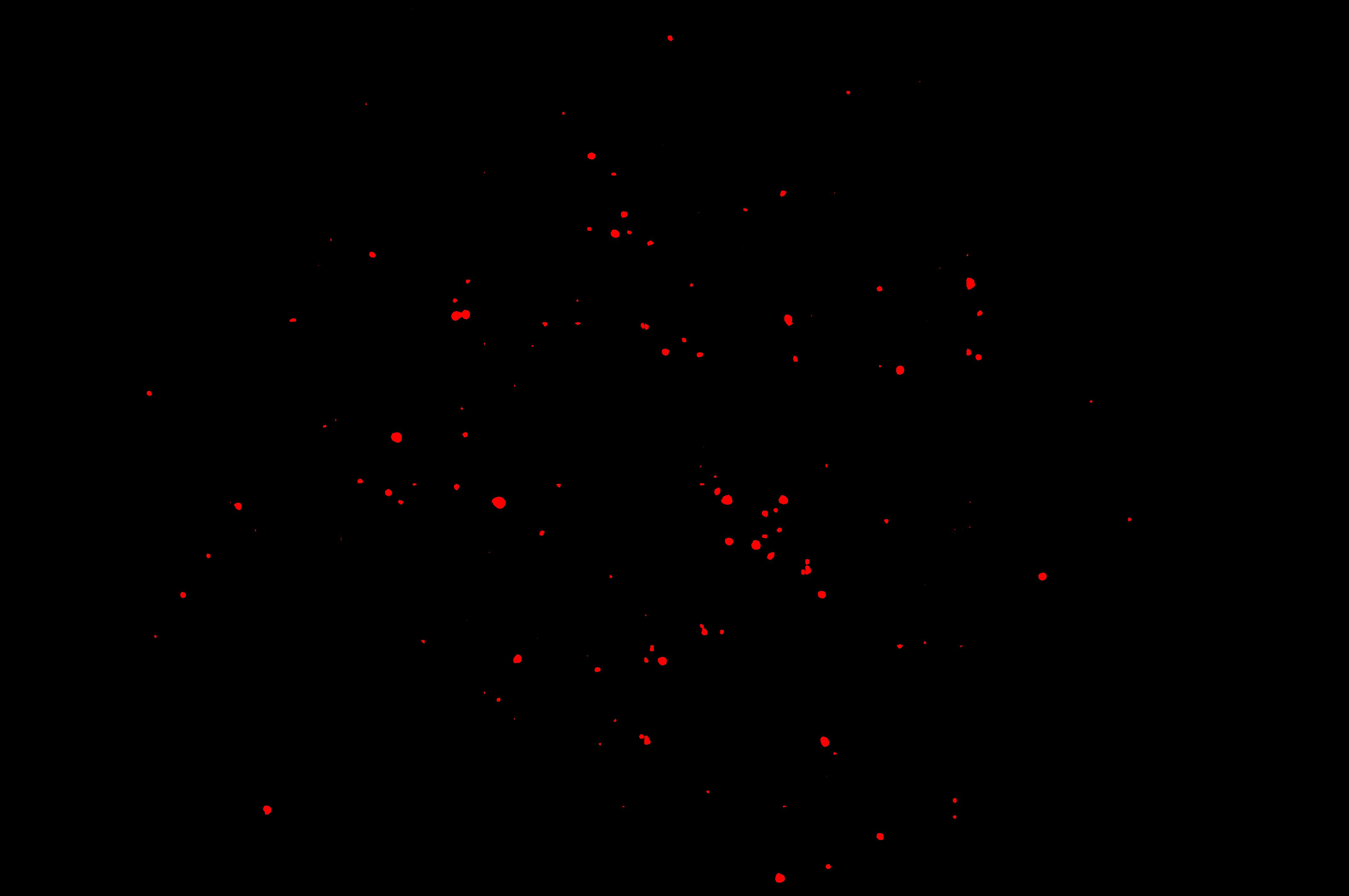}
\includegraphics[scale=0.02]{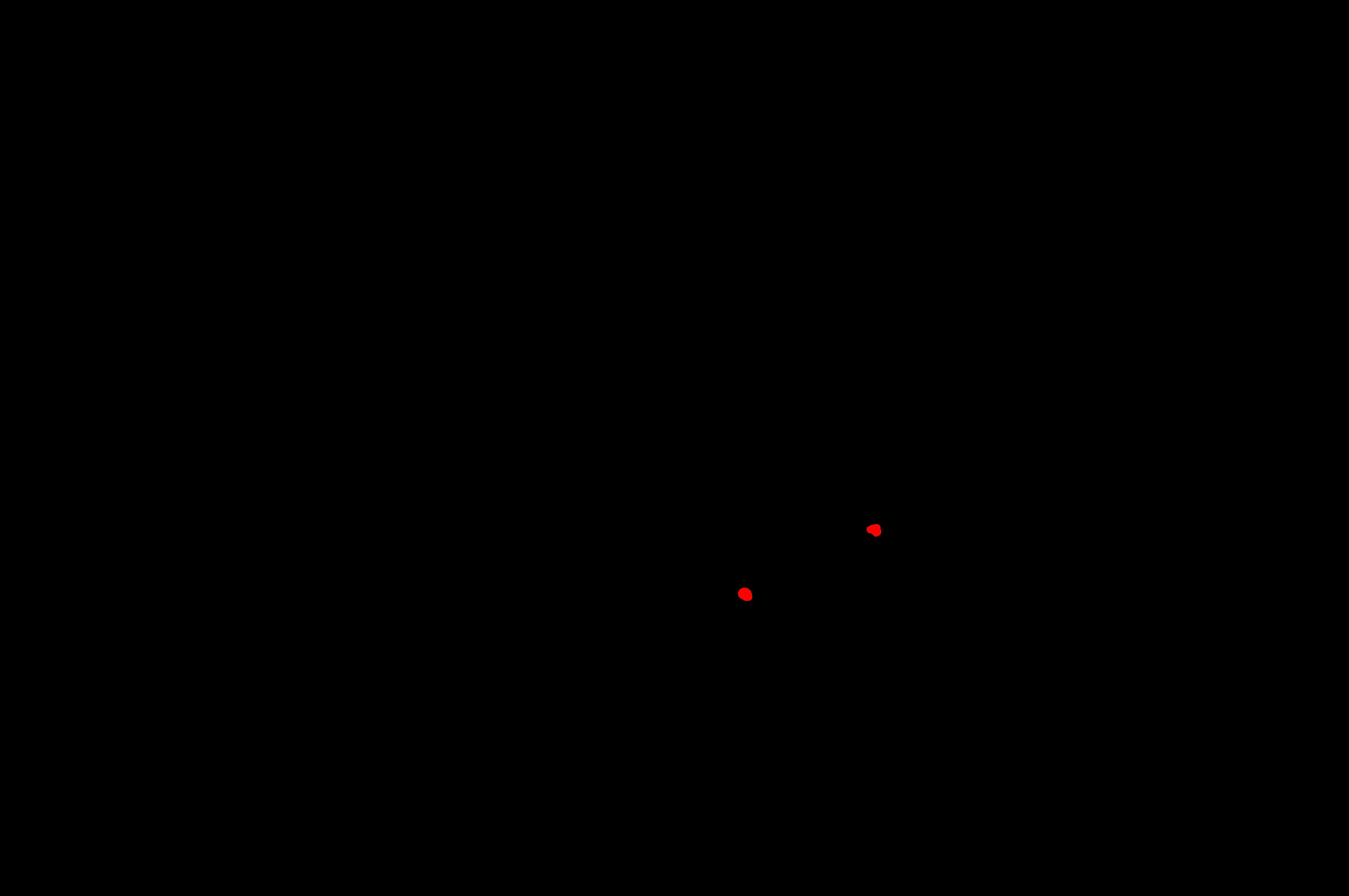}
\includegraphics[scale=0.02]{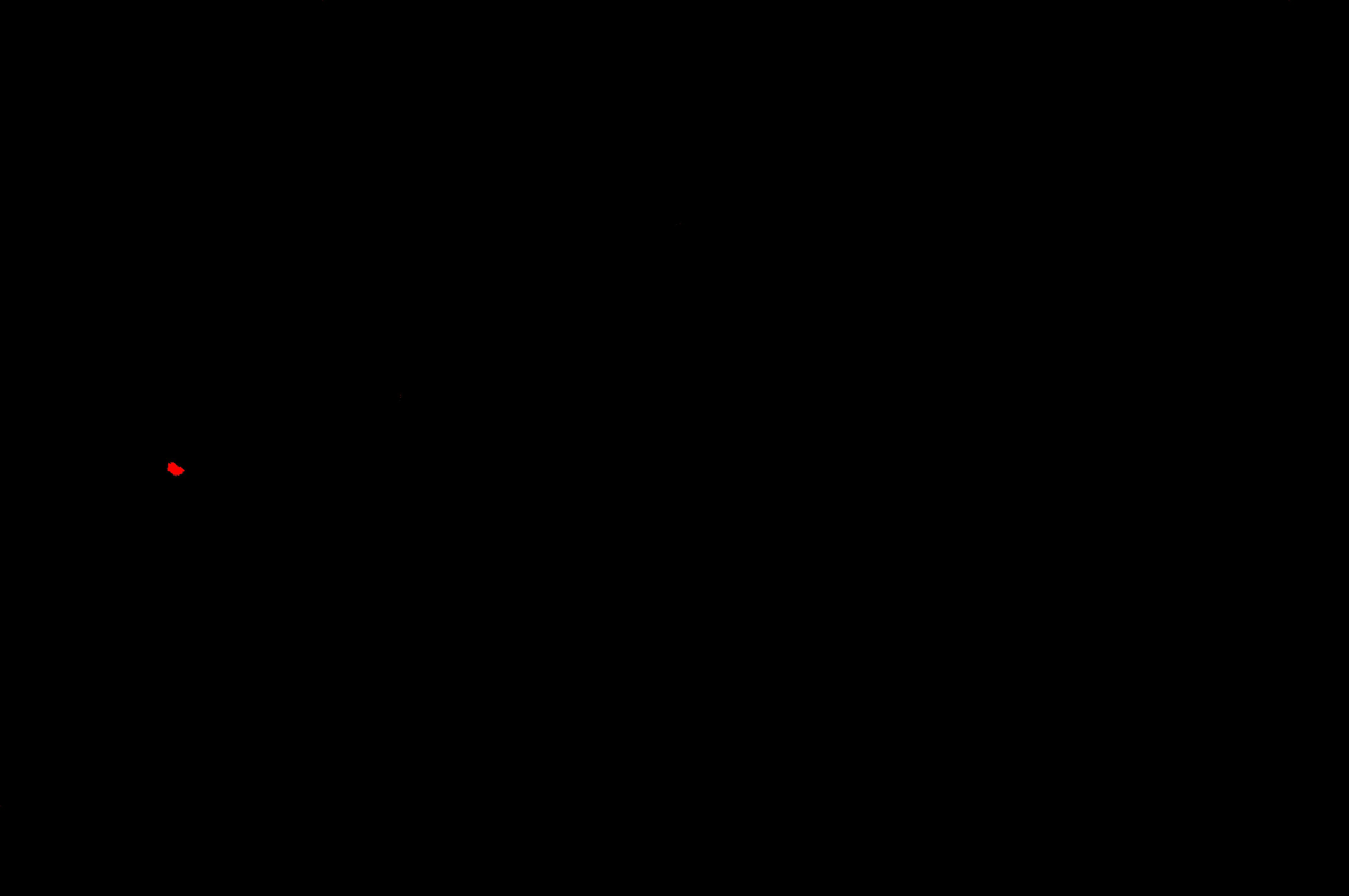}
\includegraphics[scale=0.02]{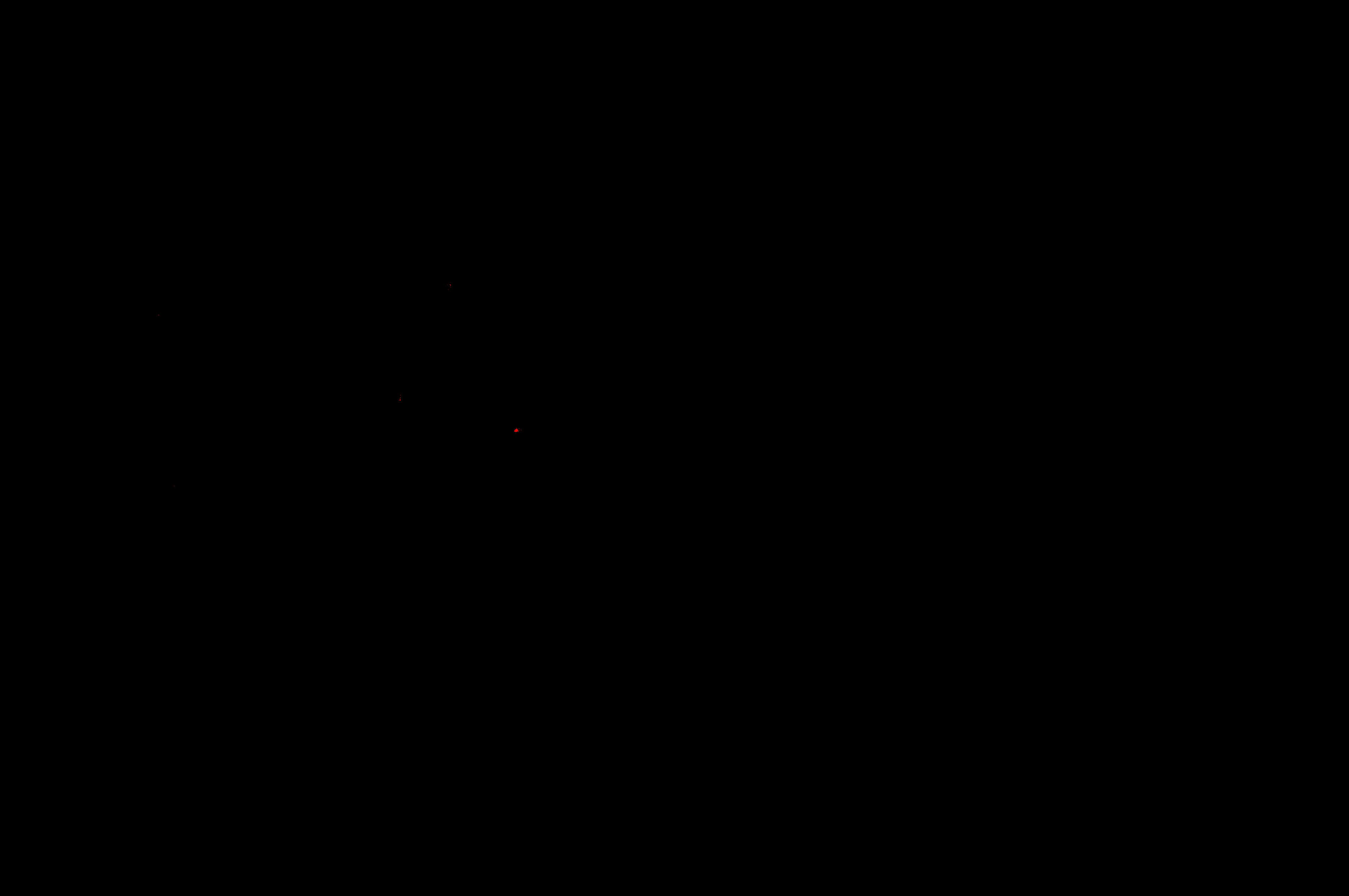}
\includegraphics[scale=0.02]{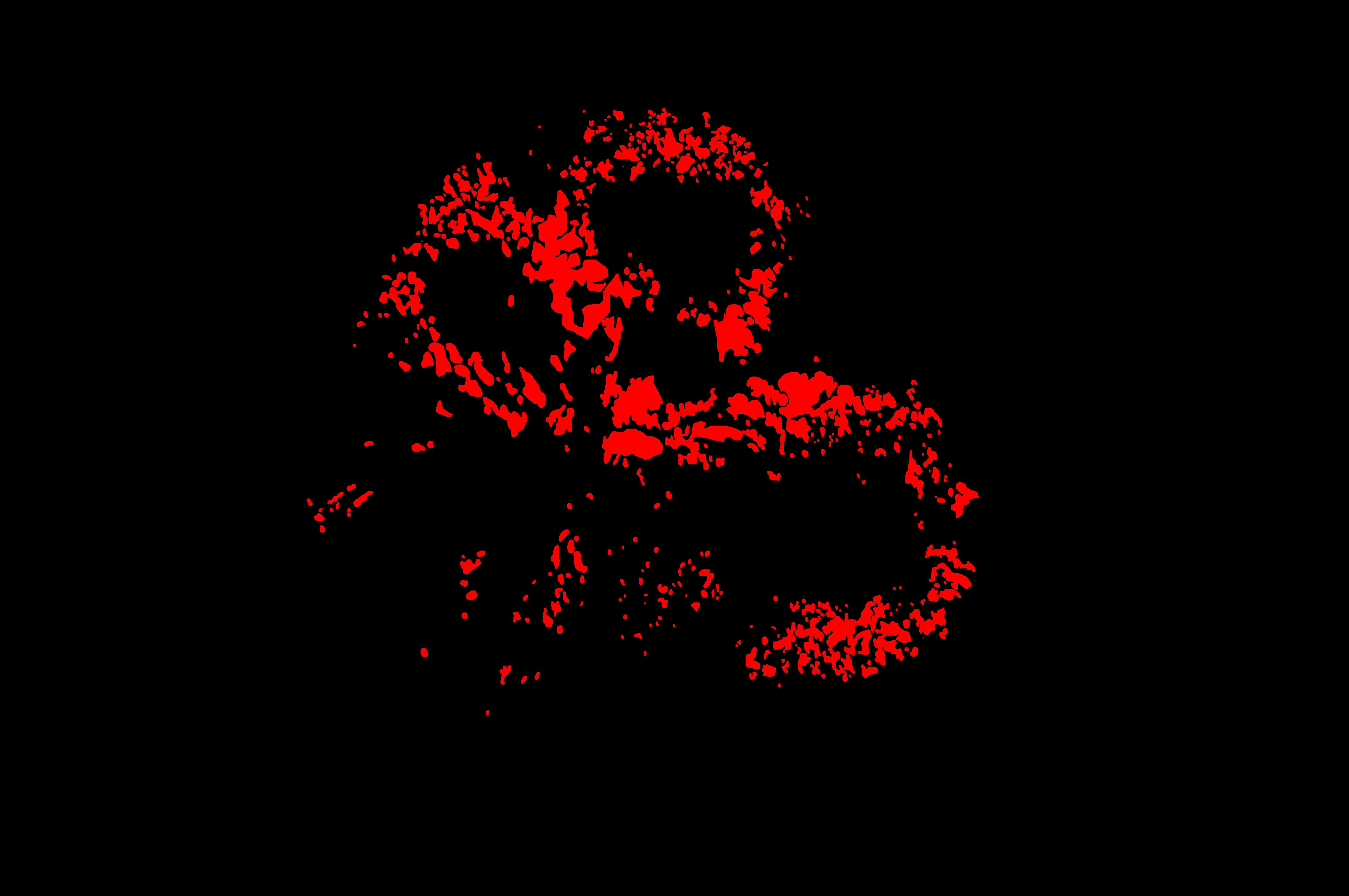}
\includegraphics[scale=0.02]{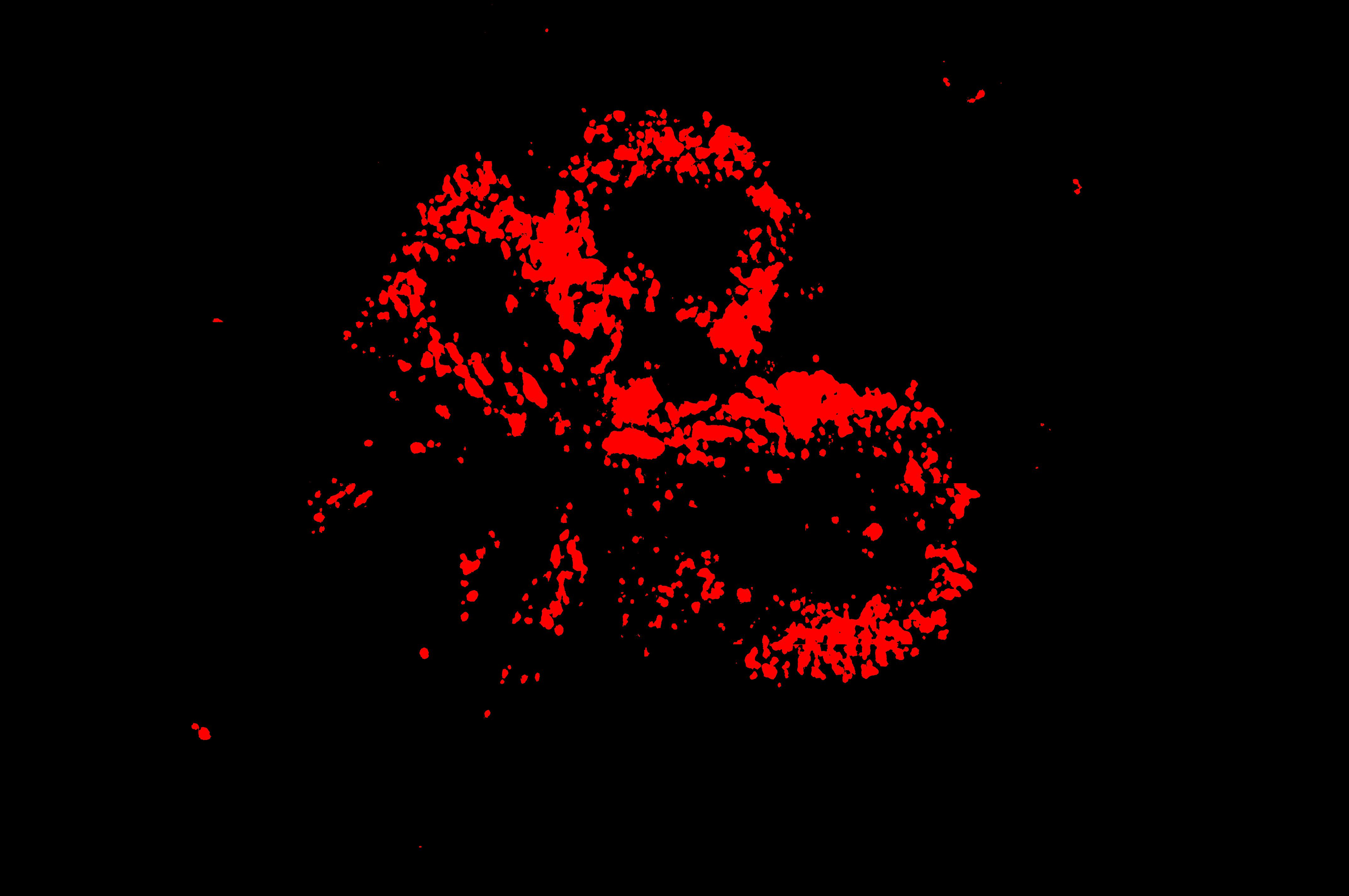}
\includegraphics[scale=0.02]{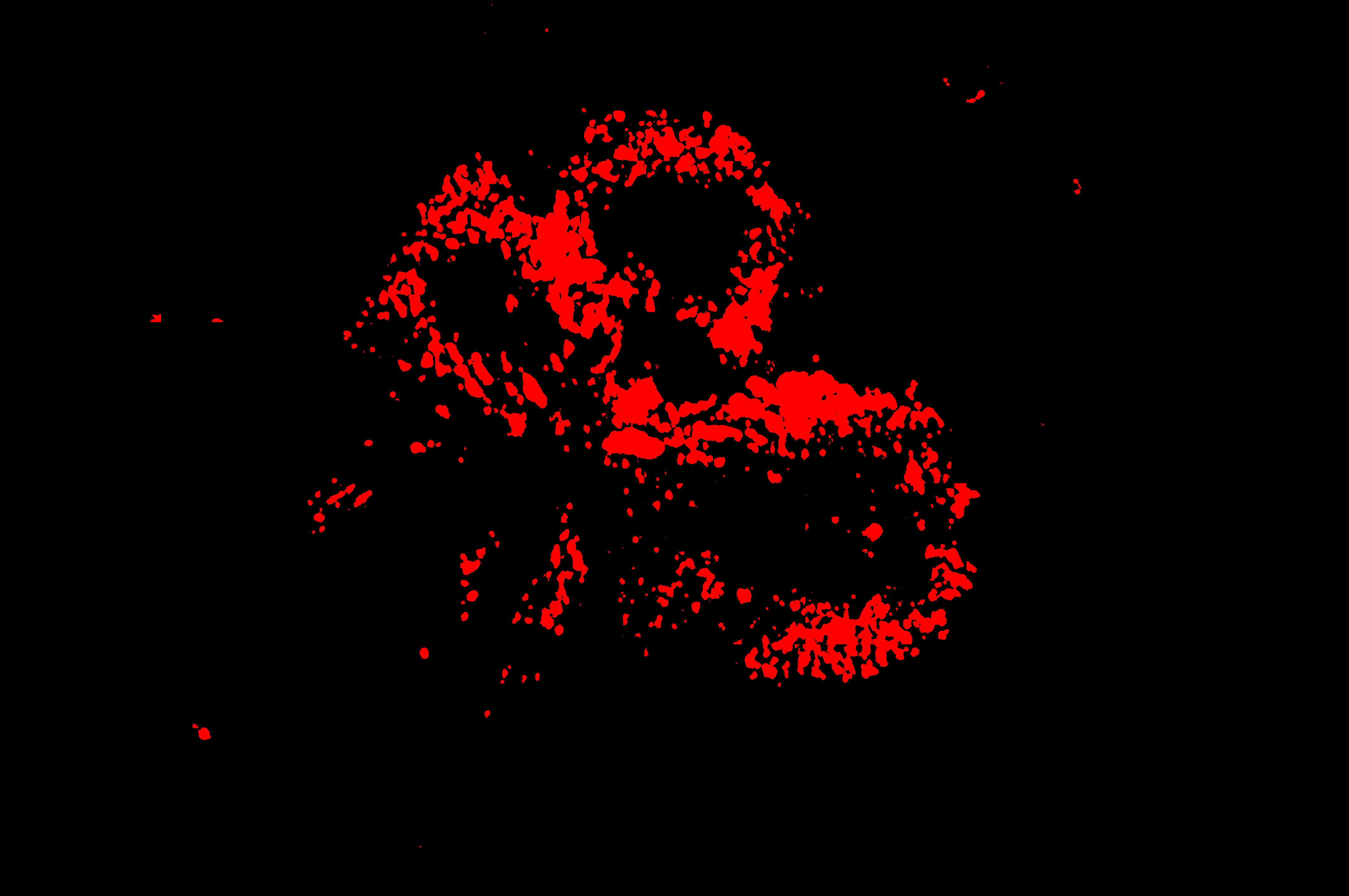}
\includegraphics[scale=0.02]{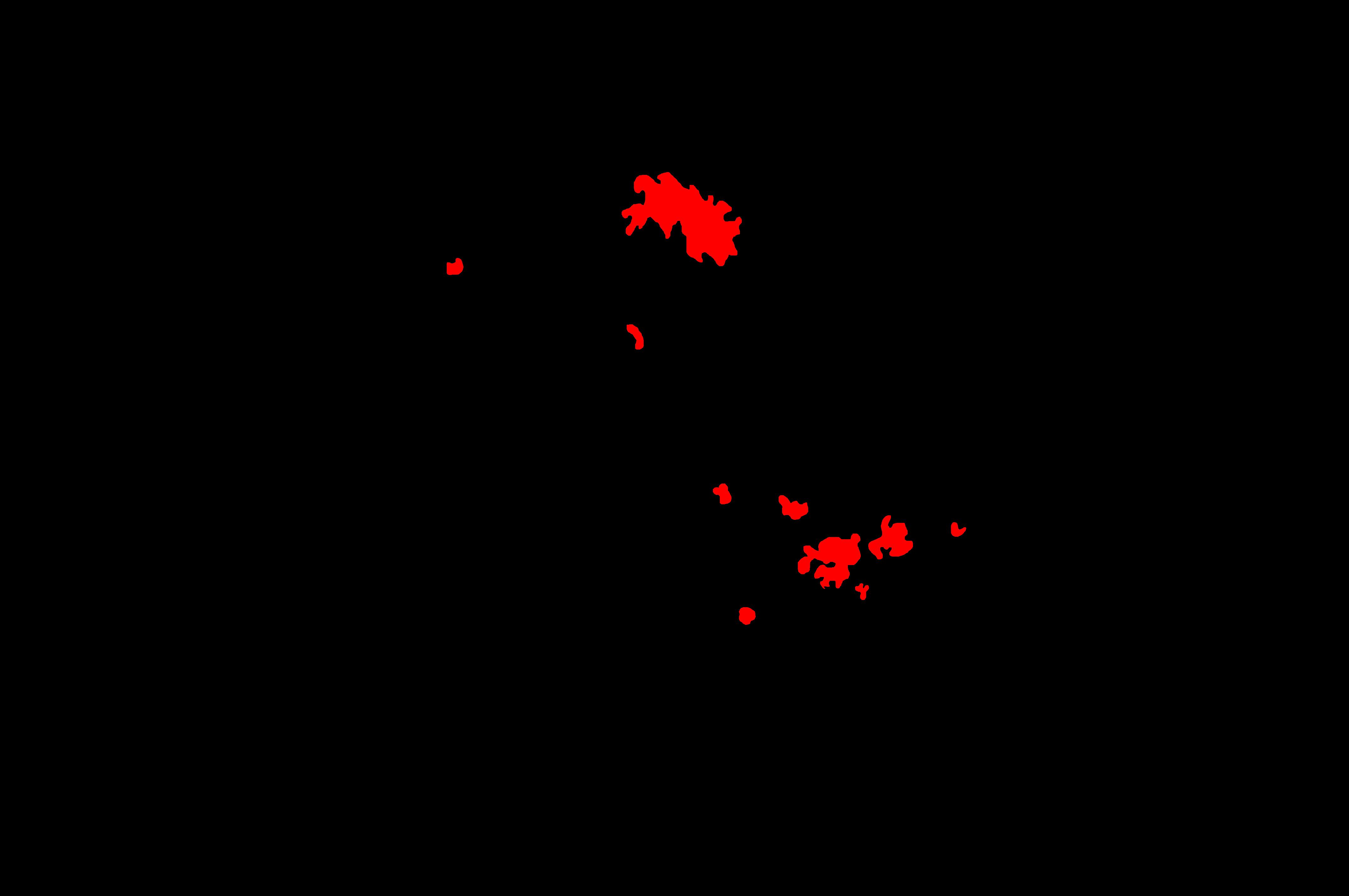}
\includegraphics[scale=0.02]{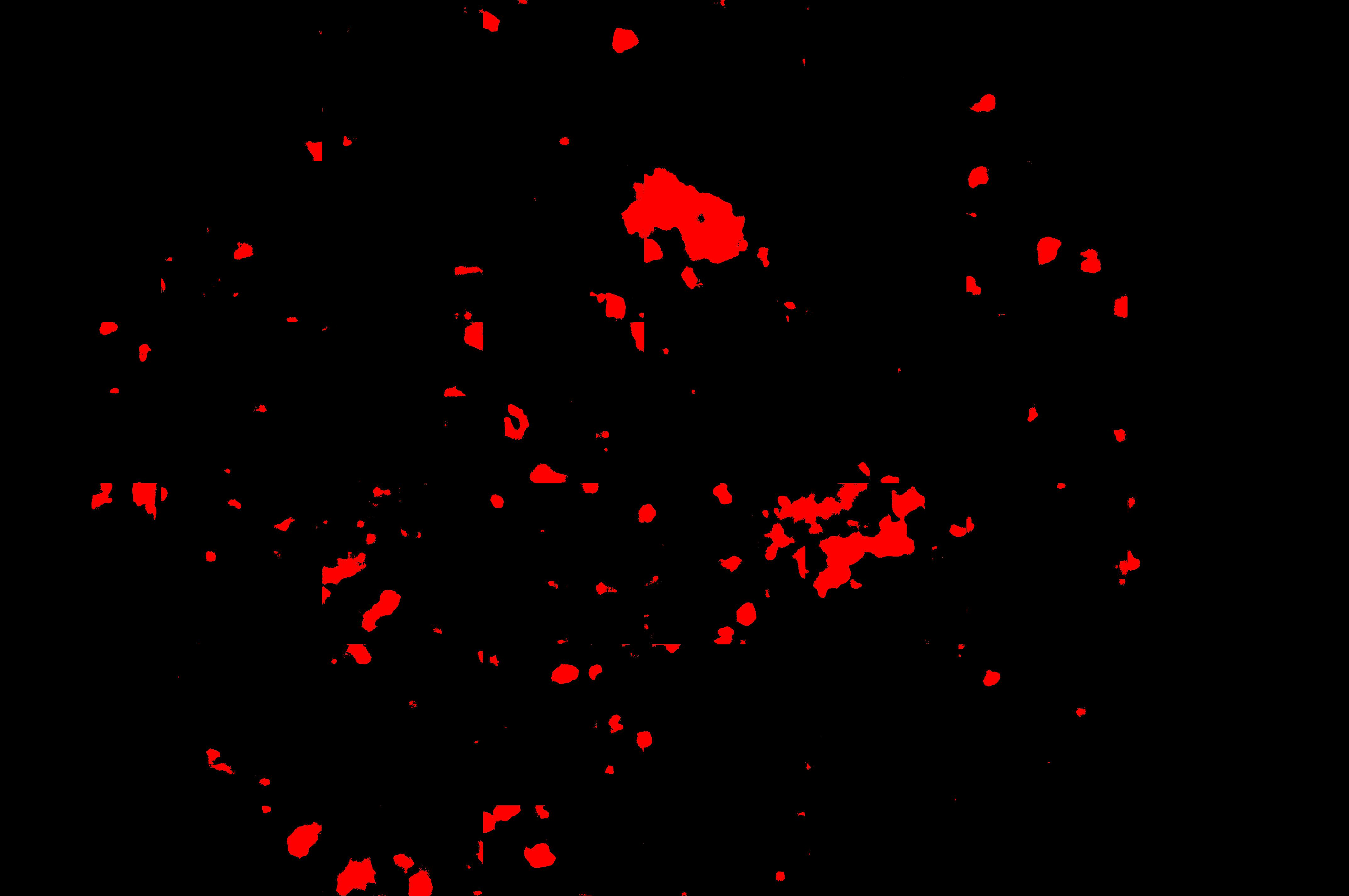}
\includegraphics[scale=0.02]{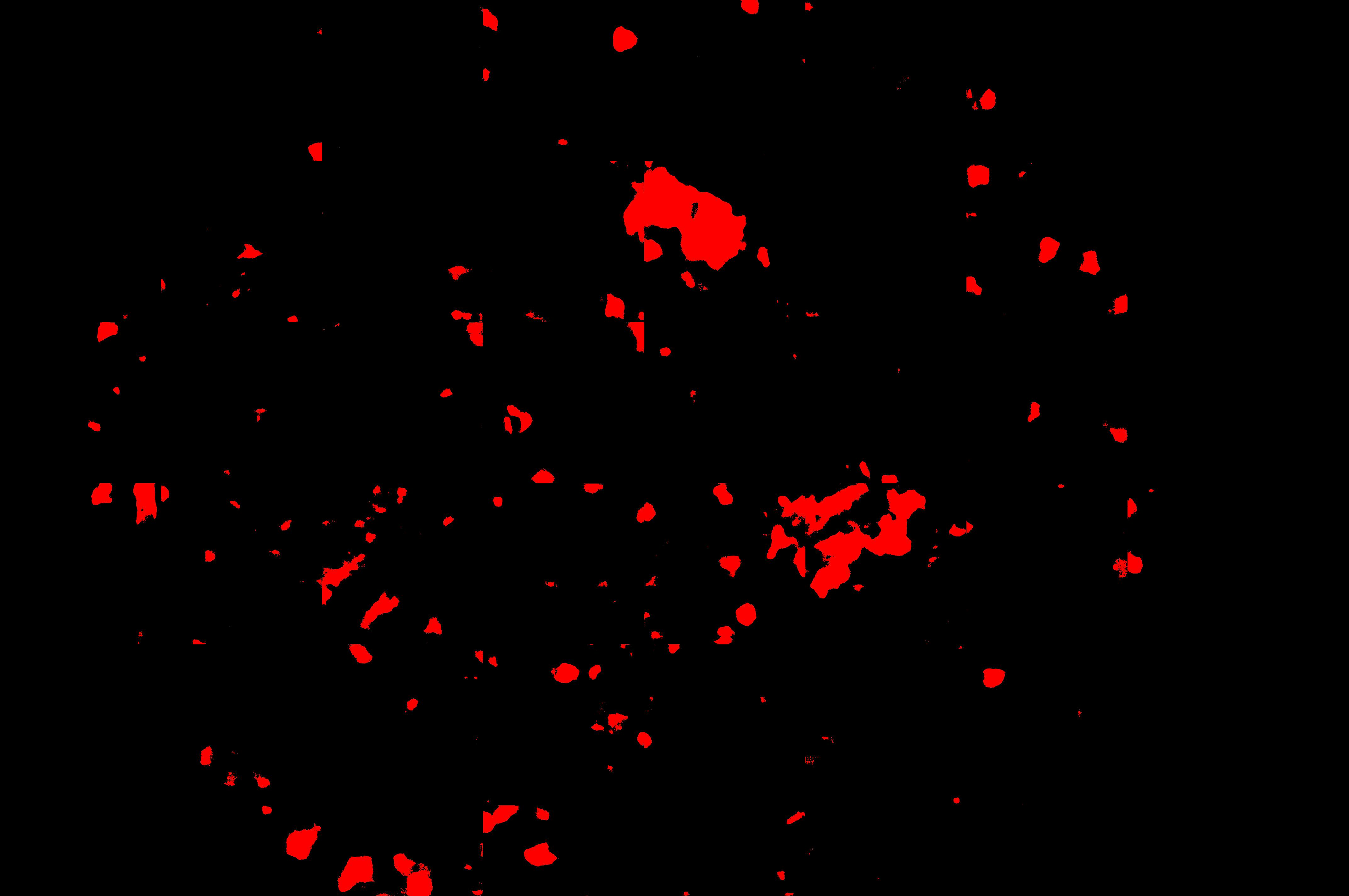}
\caption{Top: An example test set image presenting all four lesion types.  Bottom: Segmentation maps.  Each row, from top to bottom, shows lesion types: MA, SE, EX and HE.  Each column, from left to right, contains segmentation maps of ground truth, HEDNet output, and HEDNet + cGAN output, respectively.} \label{fig3}
\end{figure}

Figure~\ref{prc} presents the qualitative results of a comparison between 2 frameworks by plotting the Precision-Recall curves. The ground truth segmentation is in the second column, and the segmentation results of HEDNet and cGAN are in the last two columns. From the results, EX shows the best performance. The other 3 lesion types do not have as good results as EX, but the segmentation result of cGAN framework is much closer to the ground truth than HEDNet only, which is consistent with the quantitative results. Figure~\ref{fig3} shows an example of the original lesion image from the test set and three segmentation maps for each of the four lesion types.  

\section{Conclusion}
In this paper we have presented a method to improve the lesion segmentation performance on retinal images. We propose to use HEDNet to segment lesions in retinal images and, then, retinal image and segmentation pairs are fed to a PatchGAN discriminator that is trained to distinguish between ground truth pairs and predicted ones. The HEDNet segmentation model is then trained to both minimize a segmentation loss and to maximize the discriminator classification loss. \\

By using this approach, we show that it is possible to improve average precision on all lesion segmentation tasks. In particular, the AP of SE and HE segmentation improves by 5.3 and 3.1 percentage points when using conditional GANs over using HEDNet alone. In the future we want to evaluate if this framework is able to improve the performance in combination with other segmentation models. \\

\end{document}